\documentclass[pra,twocolumn,superscriptaddress,amsmath,amssymb,aps,floatfix]{revtex4-2}
\usepackage{dcolumn}
\usepackage{amsmath,bm,setspace}

\newcommand{\mathsym}[1]{{}}
\newcommand{\unicode}[1]{{}}
\usepackage[colorlinks=true,allcolors=blue]{hyperref}

\usepackage{graphicx}
\usepackage[toc, page]{appendix}
\usepackage{refstyle}
\usepackage{xcolor}
\usepackage{times}

\renewcommand{\thesection}{\Roman{section}}
\renewcommand\thesubsection{\Alph{subsection}}
\usepackage{titlesec}
\titleformat{\section}
{\normalfont\fontsize{9}{15}\bfseries\centering}{\thesection.}{1em}{}
\titleformat{\subsection}
{\normalfont\fontsize{9}{15}\bfseries\centering}{\thesubsection.}{1em}{}
\titlelabel{\thetitle.\quad}
\titleformat{\paragraph}
{}{\textit{(\theparagraph)}}{1em}{}
\usepackage[top=1in, bottom=1in, left=1in, right=1in]{geometry}
\usepackage{mathptmx}
\usepackage{amssymb}
\usepackage{bbold}
\usepackage{subcaption}

\usepackage{booktabs}
\usepackage{multirow}
\usepackage{adjustbox}
\usepackage{lmodern}
\usepackage{float}
\restylefloat{table}

\begin{document}
\title{Entanglement and quantum strategies reduce congestion costs in Pigou networks}
\author{Naini Dudhe}
 \email{naini.dudhe@niser.ac.in}
\author{Colin  Benjamin}
 \email{colin.nano@gmail.com}
\affiliation{School of Physical Sciences, National Institute of Science Education and Research, HBNI, Jatni 752050, India}
\begin{abstract}
Pigou's problem has many applications in real life scenarios like traffic networks, graph theory, data transfer in internet networks, etc. The two player classical Pigou's network has an unique Nash equilibrium with the Price of Stability and Price of Anarchy agreeing with each other. The situation changes for the $k$ person classical Pigou's network with $n$ being the total number of people. If we fix the behaviour of $(n-2)$ people and assume that $k$ persons take path $P_2$ where $k<(n-2)$ and the remaining take path $P_1$, the minimum cost of Nash equilibrium becomes $k$ dependent and we find a particular $k$ for which the cost is an absolute minimum. In contrast to the two person classical Pigou's network, the quantum two qubit Pigou's network with maximal entanglement gives a lower cost for the Nash equilibrium. In contrast to $k$ person classical Pigou's network, its quantum version with the quantum miracle move strategy $M$ has reduced cost for the Nash equilibrium. This has major implications for information transfer in both classical as well as quantum data networks. By employing entanglement and quantum strategies, one can significantly reduce congestion costs in quantum data networks.
\end{abstract}
  \maketitle
\section{\label{sec:level1}Introduction}
\begin{figure}[t]
\includegraphics[width=0.35\textwidth]{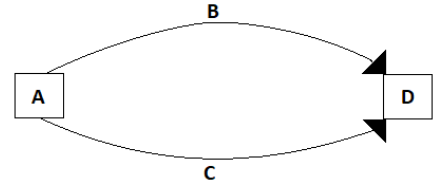}
\caption{Pigou's Two Road Problem}
\label{fig:f1}
\end{figure}

Arthur C. Pigou, a British economist, has contributed significantly to the field of welfare economics. In his book ``The Economics of Welfare"~\cite{pigou1920}, Pigou expanded on the concept of externalities, resulting from the actions of the first party which may affect unrelated third parties positively or negatively but are not included when calculating the prices.\\
Congestion externality has a space reserved as one of the most important subjects in economics. Pigou addressed this issue in his book which is well known as the two road problem~\cite{pigou1920}. In this problem, we have two roads $ABD$ and $ACD$ where $A$ is the origin and $D$ is the destination (Fig.~\ref{fig:f1}). If both the roads are identical, then traffic would be equally distributed in both. But Pigou added some constraints to this situation, that being the shifting of a few vehicles from road $B$ to road $C$ such that the trouble of driving is greatly reduced in $B$ while only slightly increasing in $C$. He proposed that a proper taxation method taking into account these circumstances will solve the problem of congestion.\\
Through this example, Pigou introduced the concept of Pigouvian taxes which now carries a significant importance in real life scenarios.\\
Traffic congestion is a major problem in all developing countries, especially China~\cite{china} due to economic and social transformation resulting in increase in vehicle ownership. This along with the urbanization has led to the decay in road infrastructure and complications for public transport. As a result, traffic congestion has affected the lives of the residents and their employment. To deal with this, the Chinese government employed several policies which ended up in failure due to inability of getting the price right for road use. After that, the government decided to introduce a plan for road pricing after taking feedback from the public. The new policy used the principle of Pigouvian taxes and congestion charges depending on the time and place of travelling. This resulted in increase in traffic speed and annual welfare gain, see Ref.~\cite{china}.\\
The impact of Pigou's problem is implicit to graph theory as it is a widely used example to depict traffic road congestion and routing games~\cite{tardos} which is a very broad topic in graph theory. Further, solutions to Pigou's problem provide assistance in routing data in large communication networks without a central authority, the internet being such an example. Pigou's problem, although was originally meant to depict congestion externality in economics, is now widely used in graph/network theory to explain data routing. A whole new subject has been formed taking Pigou's problem as the starting point known as \textit{congestion games}, {see Ref.~\cite{tardos}}\\
Classical Pigou's problem entails using classical strategies for dealing with congestion in a transport network. In the quantum version, one uses quantum strategies and/or entanglement to deal with congestion in data networks and compare these results with those obtained classically.\\
Congestion in data networks~\cite{causes_data_congestion} is analogous to traffic congestion in a populous city, the `road' being the bandwidth of the network and the `vehicle' being data being sent across the network. Data networks have information encoded in bits which traverse from origin to destination through multiple nodes. Data congestion occurs when the amount of data being sent across the network exceeds the bandwidth capacity, resulting in slow connections or loss of data packets. It can be improved by increasing the bandwidth, prioritising the traffic, using redundancy, changing the hardware to a more compatible one, eliminating bandwidth hogs, improving the network design, etc~\cite{dealing_with_congestion}. {Congestion games are an integral part of the game theory of networks, see Ref.~\cite{infofusion}. In this paper, however}, we introduce a game based on Pigou's problem to reduce data congestion using quantum strategies and entanglement without increasing the capacity of the network. We solve the game both classically and employing quantum strategies and entanglement to observe which gives a better result.\\
In the coming sections, we solve Pigou's problem classically for $2$ person and \textit{k} person cases. We will then move on to solve Pigou's problem using quantum strategies and/or entanglement for the same. We relate it to data networks and show that quantum strategies and entanglement reduces congestion costs in data networks in the discussion section. We provide two tables comparing the outcomes of classical and quantum, two person and $k$ person Pigou's network. We conclude our work by dwelling on the outcomes of the cases considered and giving a perspective on future endeavors. We finally also provide in the appendix, codes, which the reader will find useful when doing calculations.
\section{THEORY}
\subsection{Classical Pigou's network}
\begin{figure}[t]
\includegraphics[width=0.35\textwidth]{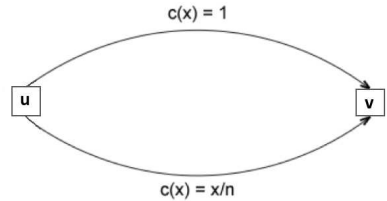}
\caption{Pigou's example - Congestion problem in a transport network}
\label{fig:f2}
\end{figure}

We start with an introduction to classical Pigou's network. The game is defined as follows:\\
Consider the simple network shown in Fig.~\ref{fig:f2}. Two disjoint edges connect a source vertex $u$ to a destination vertex $v$. Each edge is labelled with a cost function $c(x)$, which describes the cost incurred by users of that edge. The upper edge has the constant cost function $c(x)=1$ and thus represents a route that is relatively long but immune to congestion. The cost of the lower edge, given by function $c(x)=x/n$, increases as edge gets more congested where $x$ is the traffic taking that path and $n$ is the total number of persons moving from $u$ to $v$.\\
\vspace{-0.2cm}
\subsubsection{\textit{Two person case}}
Here, we have two persons (Alice and Bob) who wish to go from source node $u$ to destination node $v$. $P_1$ refers to the path with the constant cost i.e., 1 and $P_2$ refers to the path where cost varies with the amount of traffic. The cost matrix is given in Table~\ref{tab:t1}.

\begin{table}[h]
\centering
\begin{tabular}{ c  c  c }
\toprule\toprule
\caption{Cost matrix for two person case classical Pigou's game}
\label{tab:t1}
\textbf{}      & \textbf{$\ \ \ P_1$}                      & \textbf{$\ \ \ P_2$}                      \\
\midrule
\textbf{$P_1$} & \ \ \ (1, 1)                              & \ \ \ (1, $\frac{1}{2}$) \\
\midrule
\textbf{$P_2$} & \ \ \ ($\frac{1}{2}$, 1) & \ \ \ (1, 1)                 \\
\bottomrule\bottomrule     
\end{tabular}
\end{table}
The first entry of the matrix element is for Alice while the second entry is for Bob. When both Alice and Bob take path $P_1$, the cost incurred by both is 1 which corresponds to the matrix element ($P_1$, $P_1$). Similarly, when Alice takes $P_1$ and Bob takes $P_2$, Alice incurs cost 1 while Bob incurs cost 1/2 corresponding to the element ($P_1$, $P_2$) and vice versa for the case ($P_2$, $P_1$). When both Alice and Bob take path $P_2$, the cost incurred by both is 2/2 i.e., 1 which corresponds to the element ($P_2$, $P_2$).\\
\textit{\underline{The optimal solution:-}} The optimal solution is one which results in minimal total cost out of all the possible outcomes. As can be observed from the cost matrix (Table~\ref{tab:t1}), the best possible option is if one player takes $P_1$ and other takes $P_2$. So, $cost(OPT)=1+\frac{1}{2}=\frac{3}{2}$ where, $cost(OPT)$ refers to the optimal cost i.e., the total cost incurred in the optimal solution.\\
\textit{\underline{Nash equilibrium:-}} As the cost for path $P_2$ is less than or equal to 1, $P_2$ becomes a dominant strategy resulting in both players choosing $P_2$. Hence, Nash equilibrium is ($P_2$, $P_2$). Therefore,
$cost(NE)=2$ where, $cost(NE)$ refers to the cost of Nash Equilibrium i.e., the total cost incurred in Nash equilibrium.\\ As there exists a unique Nash equilibrium, we get $PoS=PoA=\frac{2}{3/2}=\frac{4}{3}$ where $PoS$ refers to Price of Stability defined as the ratio of total cost of the best Nash equilibrium (minimal cost) to that of the optimal outcome. $PoA$ refers to Price of Anarchy defined as the ratio of total cost of the worst Nash equilibrium (maximal cost) to that of the optimal outcome.
\begin{equation}\label{poseq}
\begin{array}{ccc}
  PoS  &= &\displaystyle{\frac{\textit{cost of best Nash equilibrium}}{\textit{cost of optimal outcome}}}\\\\
   PoA  & =&\displaystyle{\frac{\textit{cost of worst Nash equilibrium}}{\textit{cost of optimal outcome}}}
\end{array}
\end{equation}
The Price of Stability ($PoS$) is pertinent for games possessing multiple Nash equilibrium in order to secure the most beneficial one. On the other hand, the Price of Anarchy ($PoA$) is essential to measure the efficiency of a Nash equilibrium in a specific game. It reveals how the efficiency of a Nash Equilibrium gets worse due to the selfish behaviour of the players. If a unique Nash equilibrium exists, the $PoS$ and $PoA$ are identical.
\subsubsection{\textit{{k person case}}}
In this case, we have $n$ persons who wish to go from source node $u$ to destination node $v$ and we fix the behaviour of $(n-2)$ persons and assume that $k$ $(k < n-2)$ persons take the path $P_2$ and remaining i.e., $(n-2-k)$ take path $P_1$ and we derive the cost matrix for the remaining two persons (let them be Alice and Bob) whose behaviour is not fixed, see Table~\ref{tab:t2}.

\begin{figure}[t]
\centering
\hspace*{-1.5 cm}\includegraphics[width=0.55\textwidth]{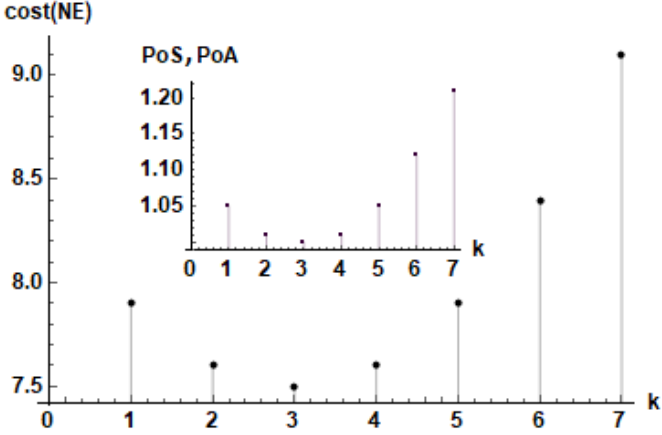}
\caption{$cost(NE)$ and $PoS$, $PoA$ versus $k$. The $k=3$ case gives us the best outcome with $cost(OPT)=cost(NE)$ and $PoS=PoA=1$}
\label{fig:f3}
\end{figure}

\begin{table}[H]
\centering
\begin{tabular}{ c  c  c }
\toprule\toprule
\caption{Cost matrix for $k$ person case}
\label{tab:t2}
\textbf{}      & \textbf{$\ \ \ P_1$}                      & \textbf{$\ \ \ P_2$}                      \\
\midrule
\textbf{$P_1$} & \ \ \ (1, 1)                              & \ \ \ (1, $\frac{k+1}{n}$) \\
\midrule
\textbf{$P_2$} & \ \ \ ($\frac{k+1}{n}$, 1) & \ \ \ ($\frac{k+2}{n}$, $\frac{k+2}{n}$)                 \\
\bottomrule\bottomrule     
\end{tabular}
\end{table}
When both Alice and Bob take path $P_1$, the cost incurred by both is 1 which corresponds to the matrix element ($P_1$, $P_1$). Similarly, when Alice takes $P_1$ and Bob takes $P_2$, Alice incurs cost 1 while Bob incurs cost $(k+1)/n$ as $k$ persons are already taking path $P_2$. This corresponds to the element ($P_1$, $P_2$) and vice versa for the case ($P_2$, $P_1$). When both Alice and Bob take path $P_2$, the cost incurred by both is $(k+2)/n$ which corresponds to the element ($P_2$, $P_2$).\\
\textit{\underline{The optimal solution:-}} As mentioned in the two person case, the optimal solution is the solution which gives us the minimum total cost (not necessarily Nash equilibrium).\\
Let us assume that $p$ persons use $P_1$ and $(n-p)$ use $P_2$. The total cost in this case is:
$$cost(OPT)=p\times1+(n-p)\times\frac{n-p}{n}$$
Our goal is to minimise the total cost. On differentiating w.r.t $p$, we get the equation
$$1+2\times\frac{(n-p)}{n}\times(-1)=0$$
This gives us the value $p=n/2$. Also $\displaystyle{\frac{d^2(cost(OPT))}{dp^2}>0}$ which tells us that the minimal cost occurs when the traffic is divided equally between the two paths.\\
Hence,
$$cost(OPT)=\frac{n}{2}\times1+\frac{n}{2}\times\frac{n/2}{n}=\frac{3n}{4}$$
Here, we take the optimal solution as the outcome which gives us the minimum cost, including all the players and considering whole range of $k$.\\
\textit{\underline{Nash equilibrium:-}} Looking at the cost matrix for the $k$ person Pigou's game, we can see that the Nash equilibrium is ($P_2$, $P_2$) since by the definition of the problem, $k+2<n$.\\
As now $(k+2)$ people choose $P_2$ {(implying $(n-k-2)$ people choose $P_1$)}, the total cost (added cost of all players) in this case then becomes
$$cost(NE)=(k+2)\times\frac{k+2}{n}+(n-k-2)\times 1$$
Hence,
\begin{equation}\label{eq:1}
cost(NE)=\frac{{(k+2)}^2}{n}+(n-k-2)
\end{equation}
Also, for $PoS$ and $PoA$, we have
{$$PoS=PoA=\frac{(k+2)\times\frac{k+2}{n}+(n-k-2)\times 1}{3n/4}$$
as there exists only one Nash equilibrium. Simplifying the expression, we get}
\begin{equation}\label{eq:2}
    PoS=PoA=\frac{4[k^2-(n-4)k+n^2-2n+4]}{3n^2}
\end{equation}
Let us fix $n=10$. In this case, the optimal solution would be the case where 5 persons take $P_1$ and 5 take $P_2$. Since, the two free players i.e., Alice and Bob always choose $P_2$, we will achieve the optimal solution identical to Nash equilibrium for the $k=3$ case, making it so that $cost(NE)=cost(OPT)=7.5$. The behaviour of $cost(NE)$ and $PoS$, $PoA$ for different values of $k$ is given in Fig.~\ref{fig:f3}.
As can be seen from Fig.~\ref{fig:f3} and equations~(\ref{eq:1}) and (\ref{eq:2}), the $PoS$, $PoA$ and $cost(NE)$ are minimum at $k=3$. So, the total number of people taking $P_2$ becomes 5 which is the case of optimal solution. Due to this, $PoS=PoA$ becomes 1 and $cost(NE)=7.5$ which is the same as $cost(OPT)$.\\
In the next section, we will introduce entanglement and quantum strategies in various cases of Pigou's Example and see if we get any better results.
\subsection{Quantum Pigou's network}
Quantum Pigou's network is obtained when we introduce quantum strategies and entanglement in classical Pigou's network. We will approach this problem using the EWL protocol~\cite{ewl}.\\
Suppose we have two players (let us call them Alice and Bob). We assign each player one qubit and the two qubit state is defined as
$|\psi\rangle=|A\rangle\otimes|B\rangle=|AB\rangle$
where, $|A\rangle=|B\rangle=|0\rangle=\left( \begin{array}{cc}
1 \\
0
\end{array} \right)$
represents the initial state of the two qubits. The qubits are entangled using the entangling gate $J=exp(-i\gamma{P_2}\otimes{P_2})$, where $\gamma\in[0,\pi/2]$ and $P_2=i\sigma_y$, the Pauli matrix. $\gamma=\pi/2$ defines maximal entanglement. Thus, the entangled two qubit state is denoted as $|\psi_0\rangle=J|AB\rangle$. The players have access to strategies $U_A$ and $U_B$ represented by:
\begin{equation}\label{eq:3}
U_i(\theta_i, \phi_i)=\left( \begin{array}{cc}
e^{i\phi_i}\cos\theta_i/2 & \sin\theta_i/2 \\
-\sin\theta_i/2 & e^{-i\phi_i}\cos\theta_i/2
\end{array} \right)
\end{equation}
where $i\in\{A,B\}$, $0\leq\theta_i\leq\pi$ and $0\leq\phi_i\leq\pi/2$. The strategy associated with path $P_1$ is given by
\begin{equation}\label{eq:4}
    P_1=U_i(0, 0)=  \mathbb{1}
\end{equation}
where $\mathbb{1}$ is the identity matrix, while the strategy associated with path $P_2$ is given by
\begin{equation}\label{eq:5}
    P_2=U_i(\pi, 0)=i\sigma_y
\end{equation}
where $\sigma_y$ is the Pauli matrix, and the quantum strategy $Q$ is denoted by
\begin{equation}\label{eq:6}
    Q=U_i(0, \pi/2)=i\sigma_z
\end{equation}
where $\sigma_z$ is the Pauli matrix. In addition to the strategies used in~\cite{ewl}, we use another strategy called the ``miracle move" strategy introduced in~\cite{comment_ewl} given by
\begin{equation}\label{eq:7}
    M=U_i(\pi/2, \pi/2)=\frac{i}{\sqrt{2}}(\sigma_y+\sigma_z).
\end{equation}
After the two players execute their moves, the state of the two qubits is $(U_A\otimes U_B) J|AB\rangle$. Finally, the disentangling operator ${J}^{\dag}$ is applied to this two qubit state to get
$|\psi_f\rangle={J}^{\dag}(U_A\otimes U_B)J|AB\rangle$.
The costs associated with both paths remain the same as in the classical case giving us the expected cost of Alice as
\begin{equation}\label{eq:8}
\$_A=1\times P_{00}+1\times P_{11}+\frac{1}{2}\times P_{10}+1\times P_{01},
\end{equation}
where $P_{\tau \tau^{\prime}}={|\langle \tau \tau^{\prime}|\psi_f\rangle|}^2$ is the probability of Alice and Bob choosing the strategies $\tau$ and $\tau^{\prime}$ respectively, $\tau,\tau^{\prime}\in\{0,1\}$. Similarly, for Bob, we get
\begin{equation}\label{eq:9}
    \$_B=1\times P_{00}+1\times P_{11}+\frac{1}{2}\times P_{01}+1\times P_{10}.
\end{equation}

\subsubsection{{\textit{Two qubit quantum Pigou's network}}}
The situation is the same as the classical two person Pigou's network. The only difference is now the two qubits are maximally entangled. The costs for the paths $P_1$ and $P_2$ remain the same as classical one. The cost matrix for this two qubit case is derived using the EWL protocol, following Eq.s~(\ref{eq:3}-\ref{eq:9}) and given in Table~\ref{tab:t3} for strategies $P_1$, $P_2$, $Q$. Here, the row player is Alice and column player is Bob. The calculation of cost is included in appendix as a mathematica notebook.\\
{\textit{Strategies $P_1$, $P_2$, $Q$:}}\\\\
The cost matrix for the entangled players, Alice and Bob, is given in Table~\ref{tab:t3}.

\begin{table}[H]
\centering
\begin{tabular}{ c  c  c  c }
\toprule\toprule
\caption{Cost matrix for maximally entangled two qubit quantum Pigou network (strategy $Q$)}
\label{tab:t3}
              & \textbf{$\ \ \ P_1$}                          & \textbf{$\ \ \ P_2$}                          & $\ \ \ Q$                         \\ \midrule
              \textbf{$P_1$}  & \ \ \ (1,   1)                            & \ \ \ (1,   0.5) & \ \ \ (1,   1)                            \\\midrule
\textbf{$P_2$}  & \ \ \ (0.5,   1) & \ \ \ (1,1)                               & (1, 0.5) \\\midrule
Q & \ \ \ (1,   1)                            & \ \ \ (0.5, 1) & \ \ \ (1,   1)      \\
\bottomrule\bottomrule          
\end{tabular}
\end{table}
From the cost matrix, we determine a unique Nash equilibrium strategy i.e., ($Q$, $Q$) with the two qubit cost, $cost(NE)=2$. However, from Table~\ref{tab:t3} it is also evident that there are four optimal solutions which are ($P_1$, $P_2$), ($P_2$, $P_1$), ($P_2$, $Q$) and ($Q$, $P_2$) with the two qubit optimal cost i.e., $cost(OPT)=3/2$. Thus, we have $PoS=PoA=4/3.$ Here, we see that we get identical results for $cost(OPT)$, $cost(NE)$ and $PoS$, $PoA$ in comparison to classical Pigou's network.\\\\
{\textit{Strategies $P_1$, P$_2$, M:}}\\\\
The strategy $M$ is the miracle move strategy defined in Eq.~(\ref{eq:7}). We again, use the EWL protocol defined in Eq.s~(\ref{eq:3}-\ref{eq:9}) for strategies $P_1$, $P_2$, $M$ and get the cost matrix as given in Table~\ref{tab:t4}.

\begin{table}[H]
\centering
\begin{tabular}{ c  c  c  c }
\toprule\toprule
\caption{Cost matrix for maximally entangled two qubit quantum Pigou's network (strategy $M$)}
\label{tab:t4}
              & \textbf{$\ \ \ P_1$}                          & \textbf{$\ \ \ P_2$}                          & $\ \ \ M$                         \\ \midrule
\textbf{$\ \ \ P_1$}   & \ \ \ (1,   1)                            & \ \ \ (1,   0.5) & \ \ \ (1,   0.75)                            \\\midrule
\textbf{$\ \ \ P_2$}   & \ \ \ (0.5,   1) & \ \ \ (1,1)                               &\ \ \ \begin{tabular}[c]{@{}c@{}}(1, 0.75)\end{tabular} \\\midrule
\textbf{$\ \ \ M$}  & \ \ \ (0.75,   1)                            & \ \ \ \begin{tabular}[c]{@{}c@{}}(0.75, 1)\end{tabular} & \ \ \ (7/8,   7/8)      \\
\bottomrule\bottomrule        
\end{tabular}
\end{table}
From the cost matrix, we determine that there is a unique pure strategy Nash equilibrium as ($M$, $M$) with $cost(NE) = 1.75$. There are two optimal strategies i.e., ($P_1$, $P_2$), ($P_2$, $P_1$) with $cost(OPT)=3/2$. From here, we can conclude that $PoS = PoA = 7/6$.\\
Comparing this to the two person classical Pigou's network and two qubit quantum Pigou's network with strategy $Q$, where we get $cost(NE)=2$ and $PoS=PoA=4/3$, we can see that the two qubit quantum Pigou's network with strategy $M$ gives a significant reduction in the cost and $PoS$, $PoA$ implying less congestion.
\subsubsection{\textit{k qubit quantum Pigou's network}}
The situation is same as the classical $k$ person Pigou's network, the only difference being that Alice and Bob, whose behaviour is not fixed and were unentangled in classical case, are now maximally entangled in quantum case. We use the EWL protocol with Alice and Bob entangled, rest $(n-2)$ are unentangled. The costs for the paths $P_1$ and $P_2$ remain the same as classical one. We fix the behaviour of those $(n - 2)$ unentangled players/qubits. Now, supposing that $k$ of them (with $k < (n - 2)$) choose path $P_2$, implying the rest unentangled players choose path $P_1$. Alice and Bob can choose either $P_1$, $P_2$ or the quantum strategy $Q$. Following Eqs.~(\ref{eq:3}-\ref{eq:9}), we derive the costs associated with the paths/strategies $P_1$, $P_2$, $Q$ and $P_1$, $P_2$, $M$ as given in Tables~\ref{tab:t5} and \ref{tab:t6} respectively, for the entangled players/qubits Alice and Bob.\\\\
{\textit{Strategies $P_1$, $P_2$, $Q$:}}

\begin{figure}[H]
\hspace*{-1.5cm}\centering
\includegraphics[width=0.55\textwidth]{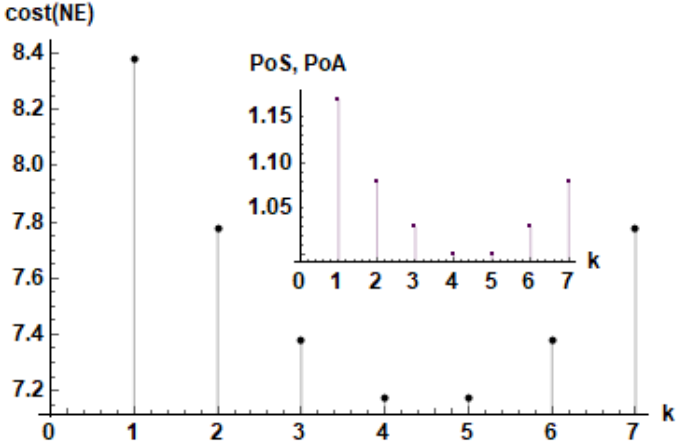}
\caption{$cost(NE)$ and $PoS$, $PoA$ in inset are plotted versus $k$. We find the best outcome at $k=4$ with $cost(OPT)=cost(NE)=7.176$ and $PoS=PoA=1$. The cost at $k=5$, although looks the same, it is actually 7.177.}
\label{fig:f6}
\end{figure}

\begin{table}[H]
\begin{adjustbox}{left}
\resizebox{\columnwidth}{!}
{
\begin{tabular}{ c  c  c  c }
\toprule \toprule
\caption{Cost matrix for $k$ qubit quantum Pigou's network with strategies/paths $P_1$, $P_2$, $Q$. The costs associated with two entangled players/qubits, Alice and Bob, are given.}
\label{tab:t5}
              & \textbf{$\ \ \ P_1$}                          & \textbf{$\ \ \ P_2$}                          & \textbf{$\ \ \ Q$}                         \\ \midrule
              
\textbf{$\ \ \ P_1$}  & \ \ \ (1,   1)                            &\ \ \  $\left(1,   \displaystyle{\frac{k+1}{n}}\right)$ & \ \ \ $\left(\displaystyle{\frac{k+2}{n}},   \displaystyle{\frac{k+2}{n}}\right)$                            \\
\midrule

\textbf{$\ \ \ P_2$}  &\ \ \ $\left(\displaystyle{\frac{k+1}{n}},   1\right)$ &\ \ \ $\left(\displaystyle{\frac{k+2}{n}},\displaystyle{\frac{k+2}{n}}\right)$                              & \ \ \ $\left(1, \displaystyle{\frac{k+1}{n}}\right)$ \\
\midrule

\textbf{$\ \ \ Q$} & \ \ \ $\left(\displaystyle{\frac{k+2}{n}},   \displaystyle{\frac{k+2}{n}}\right)$                            & \ \ \ $\left(\displaystyle{\frac{k+1}{n}}, 1\right)$ & \ \ \ (1,   1)      \\

\bottomrule\bottomrule            
\end{tabular}}
\end{adjustbox}
\end{table}
In order to compare with the classical case, let us fix $n=10$ and vary $k$ from 1 to 7. The cost matrix as given in Table~\ref{tab:t5} is for the entangled players/qubits Alice and Bob, with costs associated to each given as $\$_A$, $\$_B$ (see, Eqs.~(\ref{eq:8}), (\ref{eq:9}) for our method to calculate these costs in the quantum case) which are affected by the behaviour of the rest of the players/qubits whose behaviour is fixed. We calculate $PoS$, $PoA$, $cost(NE)$, $cost(OPT)$ and $\$_A$, $\$_B$, as shown in Table~\ref{tab:t5}. This calculation has been done in mathematica and the notebook has been appended to the appendix.\\ Similar to the classical $k$ person case, we choose the optimal solution as one which gives the minimal total cost (added cost of all players), taking into account the whole range of $k$. We see that the optimal solution is the same as the Nash equilibrium for $k=4$ case which gives the minimum cost over the whole range of $k$ with $cost(NE)=cost(OPT)=7.176$. We observe a unique mixed Nash equilibrium for each $k$. Now, computing $cost(NE)$ and $PoS$, $PoA$, we see, for $k=1$, $cost(NE)=8.38$, $PoS=PoA=1.17$. Similarly, for $k=2 \text{ and } k=7$, $cost(NE)=7.78$, $PoS=PoA=1.08$. Again, $k=3 \text{ and } k=6$, $cost(NE)=7.38$, $PoS=PoA=1.03$. Finally, for $k=4$, $cost(NE)=7.176$, $PoS=PoA=1$ and for $k=5$, $cost(NE)=7.177$ and $PoS=PoA=1.0001$. In Fig.~\ref{fig:f6} and its inset, we plot the Nash equilibrium cost, $cost(NE)$ along with $PoS$ and $PoA$ versus $k$.

In its classical counterpart (see Fig.~\ref{fig:f3}), we get $cost(OPT)$ and the minimum cost of Nash equilibrium at 7.5. On the other hand, if we compare this classical result with the quantum version, we conclude that the $k$ qubit quantum Pigou's network with strategy $Q$ gives us a significant reduction in the cost. This happens for the case $k=4$ with $cost(NE)=cost(OPT)=7.176$. It is interpreted as the case where the free players/qubits are equally divided between the paths $P_1$ and $P_2$, and Alice and Bob each take paths $P_1$ and $P_2$ both with mixed probability 4/17 and the strategy $Q$ with mixed probability 9/17. This implies significant reduction in congestion cost and signifies that in quantum data networks, entanglement along with quantum paths or quantum strategies, will lead to better results.\\ \\
{\textit{Strategies $P_1$, $P_2$, $M$:}}\\\\
Using EWL protocol and following Eqs.~(\ref{eq:3}-\ref{eq:9}), we obtain the cost matrix for the entangled players/qubits i.e., Alice and Bob for strategies $P_1$, $P_2$ and $M$ depicted in Table~\ref{tab:t6}. The costs incurred by Alice and Bob are affected by the behaviour of the rest of the players/qubits whose behaviour is fixed.

\begin{figure}[H]
\centering\includegraphics[width=0.55\textwidth]{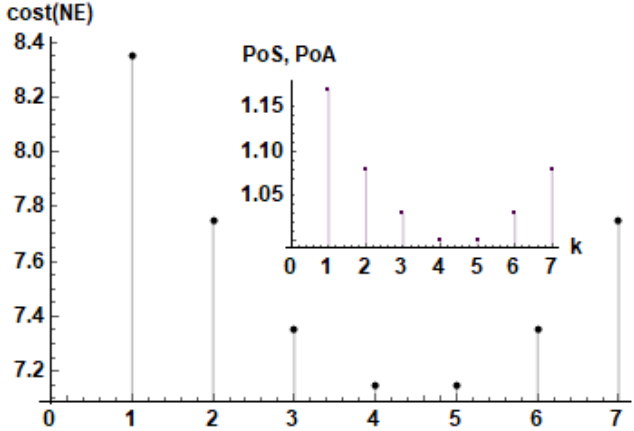}
\caption{$cost(NE)$ and $PoS$, $PoA$ in inset are plotted versus $k$. We find the best outcome at $k=4$ and $k=5$ with $cost(OPT)=cost(NE)=7.15$ and $PoS=PoA=1$.}
\label{fig:f7}
\end{figure}

\begin{table}[H]
\begin{adjustbox}{left}
\resizebox{1.2\columnwidth}{!}
{
\begin{tabular}{ c  c  c  c }
\toprule \toprule
\caption{Cost matrix for $k$ qubit quantum Pigou's network with strategies/paths $P_1$, $P_2$, $M$. The costs associated with two entangled players/qubits, Alice and Bob, are given.}
\label{tab:t6}
              & \textbf{$\ \ \ P_1$}     & \textbf{$\ \ \ P_2$}   & \textbf{$\ \ \ M$}  \\
              \midrule
\textbf{$\ \ \ P_1$}  & $\ \ \ $(1,   1) & \ \ \ $\left(1, \displaystyle{\frac{k+1}{n}}\right)$ & $\ \ \ \left(\displaystyle{\frac{n+k+2}{2n}},\displaystyle{\frac{2k+3}{2n}}\right)$    \\
        \midrule
\textbf{$\ \ \ P_2$}  & $\ \ \ \left(\displaystyle{\frac{k+1}{n}},   1\right)$ & \ \ \ $\left(\displaystyle{\frac{k+2}{n}}, \displaystyle{\frac{k+2}{n}}\right)$                               & $\ \ \ \left(\displaystyle{\frac{n+k+2}{2n}},\displaystyle{\frac{2k+3}{2n}}\right)$ \\
       \midrule
\textbf{$\ \ \ M$} & $\ \ \ \left(\displaystyle{\frac{2k+3}{2n}},   \displaystyle{\frac{n+k+2}{2n}}\right)$                            & $\ \ \ \left(\displaystyle{\frac{2k+3}{2n}},\displaystyle{\frac{n+k+2}{2n}}\right)$ & $\ \ \ \left(\displaystyle{\frac{2n+2k+3}{4n}},   \displaystyle{\frac{2n+2k+3}{4n}}\right)$      \\
\bottomrule\bottomrule
\end{tabular}}
\end{adjustbox}
\end{table}
Once again, in order to compare with its classical counterpart, we fix $n=10$ and vary $k$ from 1 to 7. The optimal solution is the one with the minimum cost out of all solutions and same as Nash equilibrium for $k=4$ and $k=5$ case with $cost(OPT)=cost(NE)=7.15$. We observe that there exists unique pure Nash equilibrium ($M$, $M$) for all $k$ in contrast to the case of $P_1$, $P_2$, $Q$ where we observed mixed Nash equilibrium for all $k$. For the $k=4$ and $k=5$ case, the entangled players, Alice and Bob both adopt the miracle move strategy $M$, giving $cost(NE)=cost(OPT)=7.15$. Now, computing $cost(NE)$ and $PoS$, $PoA$, we see, for $k=1$, $cost(NE)=8.35$, $PoS=PoA=1.17$. Similarly, for $k=2$ and $k=7$, $cost(NE)=7.75$, $PoS=PoA=1.08$. Again for $k=3$ and $k=6$, $cost(NE)=7.35$, $PoS=PoA=1.03$. Finally, for $k=4$ and $k=5$, $cost(NE)=7.15$, $PoS=PoA=1.$. The plot given in Fig.~\ref{fig:f7} and its inset depicts the behaviour of $cost(NE)$ and $PoS$, $PoA$ versus $k$.
Comparing this to the $k$ person classical Pigou’s network (where we get $cost(OPT)=cost(NE)=7.5$) and  $k$  qubit  quantum  Pigou’s  network with  strategy $Q$ (where we get $cost(OPT)=cost(NE)=7.176$), we can see that the $k$ qubit quantum Pigou's network with strategy $M$ reduces the cost a bit more (with $cost(OPT)=cost(NE)=7.15$), resulting in a reduction in congestion.\\
In the next section, we will compare our results achieved after employing quantum strategies and entanglement with those obtained classically and identify which is more feasible.
\section{DISCUSSION ON CLASSICAL VERSUS QUANTUM PIGOU'S NETWORK}
Here, we compare our own results achieved classically to those achieved after employing quantum strategies and entanglement. The comparison for two person and $k$ person case is given in Tables~\ref{tab:t7} and \ref{tab:t8} respectively. We find that quantum strategies and entanglement lead to reduced congestion costs as compared to classical version in both two and $k$ person case as shown below.

\begin{table*}[ht]
\begin{adjustbox}{center}
\resizebox{2.5\columnwidth}{!}{
\begin{tabular}{ c  c  c }
\toprule\toprule
\caption{Comparison between two person classical Pigou's network and quantum Pigou's network}
\label{tab:t7}
\multirow{3}{*}{\text{Classical Pigou's network}} & \multicolumn{2}{c}{\multirow{2}{*}{\text{Quantum Pigou's network}}} \\
 & \multicolumn{2}{c}{} \\
 \cmidrule{2-3}
 & \textbf{($P_1$, $P_2$, $Q$)} & \textbf{($P_1$, $P_2$, $M$)} \\ \midrule
\begin{tabular}[c]{@{}c@{}}The only Nash equilibrium is ($P_2$, $P_2$)\\with $cost(NE)=2$ and the optimal solution\\is ($P_1$, $P_2$) or ($P_2$, $P_1$) with $cost(OPT)=1.5$\\and $PoS=PoA=4/3.$\end{tabular} &\ \ \ \ \ \begin{tabular}[c]{@{}c@{}}This gives identical results as the classical\\case, the only difference being that the\\Nash equilibrium changes to ($Q$, $Q$) with\\$cost(NE)=2$,
$cost(OPT)=3/2$\\and $PoS=PoA=4/3$.\end{tabular} &\ \ \ \ \ \begin{tabular}[c]{@{}c@{}}This case gives us the best result\\
with ($M$, $M$) as the Nash equilibrium with\\$cost(NE)=7/4$, $cost(OPT)=3/2$ and\\$PoS=PoA=7/6$ which is better than other\\two cases.\end{tabular} \\ \bottomrule\bottomrule
\end{tabular}}
\end{adjustbox}
\end{table*}

\begin{table*}[ht]
\begin{adjustbox}{center}
\resizebox{2.5\columnwidth}{!}{
\begin{tabular}{ c  c  c }
\toprule\toprule
\caption{Comparison between $k$ person classical Pigou's network and quantum Pigou's network}
\label{tab:t8}
\multirow{3}{*}{\text{Classical Pigou's network}} & \multicolumn{2}{c}{\multirow{2}{*}{\text{Quantum Pigou's network}}} \\
 & \multicolumn{2}{c}{} \\
\cmidrule{2-3}
 & \textbf{($P_1$, $P_2$, $Q$)} & \textbf{($P_1$, $P_2$, $M$)} \\ \midrule
\begin{tabular}[c]{@{}c@{}}The optimal solution is the case\\ where the players are equally divided\\between both paths. The outcome\\gets better as $k$ nears to 3 (Fig.~\ref{fig:f3})\\so that when Alice and Bob are\\included, we get the case where 5 people\\take $P_1$ and 5 take $P_2$ as Nash equilibrium\\for Alice and Bob is ($P_2$, $P_2$). In this\\case, $cost(OPT)=cost(NE)=7.5.$\end{tabular} & \ \ \ \ \ \begin{tabular}[c]{@{}c@{}}We see that optimal solution\\is identical to Nash equilibrium\\for the $k=4$ case having an unique\\mixed Nash equilibrium with\\$cost(NE)=cost(OPT)=7.176$. The\\behaviour of $cost(NE)$ and $PoS$,\\$PoA$ is shown in Fig. \ref{fig:f6}.\end{tabular} & \ \ \ \ \ \begin{tabular}[c]{@{}c@{}}This case gives us the best possible\\ solution with the optimal solution\\obtained at $k=4$ and $k=5$ which\\is also the Nash equilibrium with\\$cost(OPT)=cost(NE)=7.15$. The Nash\\equilibrium for Alice and Bob is ($M$, $M$).\end{tabular} \\ \bottomrule\bottomrule
\end{tabular}}
\end{adjustbox}
\end{table*}
\subsection{Two person case}
For classical case, we found that the optimal solution is ($P_1$, $P_2$) or ($P_2$, $P_1$) with $cost(OPT)=3/2$ while the Nash equilibrium is found to be ($P_2$, $P_2$) with $cost(NE)=2$. This gives us $PoS=PoA=4/3$ as there exists a unique Nash equilibrium. The quantum version with strategies $P_1$, $P_2$, $Q$ gives us a result identical to classical case, the only difference being that Nash equilibrium changes to ($Q$, $Q$). The benefit of using quantum strategies and entanglement shows itself in the quantum case with strategies $P_1$, $P_2$, $M$ where optimal solution remains the same as classical but Nash equilibrium changes to ($M$, $M$) with $cost(NE)=7/4$. This results in $PoS=PoA=7/6$ which is noticeably better than both classical case and quantum case with strategies $P_1$, $P_2$, $Q$. In the language of data networks, the classical paths are the two distinct independent paths $P_1$ and $P_2$ available for data packs to move between nodes $u$, $v$ as in Fig.~\ref{fig:f2}. The quantum strategy $Q$ is associated with taking paths $P_1$ and $P_2$ with a phase and miracle move strategy $M$ entails superposition of paths $P_1$ and $P_2$. In this two person case, we found that data packs using strategy $M$ to travel between nodes $u$, $v$ will experience less congestion as compared to data packs travelling via classical paths $P_1$ and $P_2$.
\subsection{\textit{k} person case}
For classical case, we found that the optimal solution is the same as Nash equilibrium at $k=3$ with $cost(OPT)=cost(NE)=7.5$. The Nash equilibrium is found to be ($P_2$, $P_2$) for Alice and Bob and $cost(NE)$ varies with $k$, achieving minimum at $k=3$ with $cost(NE)=7.5$ (see Fig.~\ref{fig:f3}). Thus, $PoS$, $PoA$ also vary with $k$, achieving minimum at $k=3$ with $PoS=PoA=1$ as there exists a unique Nash equilibrium for each $k$. The quantum version with strategies $P_1$, $P_2$, $Q$ gives us an optimal solution identical to the mixed Nash equilibrium achieved at $k=4$ with $cost(OPT)=cost(NE)=7.176$. We see an unique mixed Nash equilibrium for each $k$. Similar to classical case, here as well, Nash equilibrium and $PoS$, $PoA$ vary with $k$. The minimum value is for $k=4$, wherein $cost(NE)=7.176$ and $PoS=PoA=1$. It can be seen that using quantum strategy $Q$ here gives us a Nash equilibrium and optimal solution better than the classical one. The result improves a little more for the case with strategies $P_1$, $P_2$, $M$ with optimal solution obtained for $k=4$ which is also the Nash equilibrium. We have $cost(OPT)=cost(NE)=7.15$ for $k=4$. For the strategies $P_1$, $P_2$, $M$, we only see an unique pure Nash equilibrium ($M$, $M$) for Alice and Bob at each $k$. The $cost(NE)$ along with $PoS$, $PoA$ reach minimum values for $k=4$ and $k=5$ for the $P_1$, $P_2$, $M$ strategies. For $k=4$ and $k=5$, we have $cost(NE)=7.15$ and $PoS=PoA=1$ for the strategies $P_1$, $P_2$, $M$. Thus, it is clear that information encoded on qubits using quantum strategies $P_1$, $P_2$, $M$ or $P_1$, $P_2$, $Q$ are subject to reduced congestion costs than that for classical bits using paths $P_1$ and $P_2$ to move from nodes $u$ to $v$. Moreover, data transfer using qubits is faster as compared to that using classical bits. {Furthermore, quantum communication is much safer as compared to classical communication, see Ref.~\cite{qkd}. As far as classical communication is concerned, sharing of data takes place through the distribution of a shared key to the sender and receiver which is used for encryption of the message. Since the receiver possesses the shared key, they can decode the data at their end. Classical communication ensures security through an algorithm for creating keys, which although difficult to break, is still vulnerable to hackers. Quantum communication, on the other hand, is more secure utilizing a concept known as \textit{quantum key distribution (QKD)}. A piece of classical data is encrypted through the key being encoded onto qubits. The qubits are then sent by the sender to the receiver, who measures the qubits to get the key values. Since QKD employs the quantum approach, it is easy to discern interference by a third party because the moment someone measures the qubits, the key collapses, changing the state of the qubits. This, along with congestion reduction achieved using our method can make data packs travelling across networks significantly easier, faster and much safer.}
\subsection{Comparison with related approaches}
In this subsection, we compare the results obtained in our paper with that obtained in Ref.~\cite{scarpa}. We state how these two works differ and how we achieve better results. Our focus in Ref.~\cite{scarpa} is on section 5 which introduces the quantum version of Pigou’s example using the EWL protocol~\cite{ewl}.\\
The network used is the same as two person Pigou's network except that it is with $n$ players. Ref.~\cite{scarpa} starts off with introducing the classical Pigou's network as is already done in our paper in section II.A. Now, in order to use the EWL protocol with $n$ players, they create ``$Entanglment\ by\ couples$" (if the number is odd, the last one is free). The behaviour of $(n-2)$ players is fixed, assuming that $k\ (k<n-2)$ choose path P$_2$, the cost matrix has been calculated for the remaining two players. In this paper, the degree of entanglement is assumed to be maximal ($\gamma = \pi/2$). After calculation, the cost matrix without the introduction of any new strategies, comes out to be the same as the classical $k$ person case. In Ref.~\cite{scarpa}, the quantum strategies introduced, which has been derived from that used in Ref.~\cite{ewl} and is mentioned below,
\begin{equation}
 \mbox{for Path 1:} \left(\begin{array}{cc}
-i & 0\\
0 & i
\end{array} \right) \mbox{ and for Path 2: }  \left( \begin{array}{cc}
0 & -i\\
-i & 0
\end{array} \right).
\end{equation}
Using these strategies, Ref.~\cite{scarpa} obtains $cost(OPT)=cost(NE)=3n/4$ and $PoS=PoA=1$. The difference in our methods and theirs is that while they have generalised the value of $k$, we have taken a case by case approach and monitored the evolution of the game. We also implement different strategies as compared to Ref.~\cite{scarpa}, i.e., the quantum strategies $Q$ and the miracle move strategy $M$. Now, in order to compare our results with Ref.~\cite{scarpa}, let us fix $n=10$. This gives us $cost(OPT)=cost(NE)=7.5$. On the other hand, for the $k=4$ case, the strategies $Q$ and $M$ give us $cost(OPT)=cost(NE)=7.176$ and $cost(OPT)=cost(NE)=7.15$ respectively which are both better than $7.5$ (the Result from Ref.~\cite{scarpa}), thus proving our results worth and uniqueness. Also, in case of Ref.~\cite{scarpa}, quantum strategies do not give rise to better results than the best classical result, i.e., the optimal solution. In our case, even the quantum optimal solution changes and gives better $cost(OPT)$ and $cost(NE)$ than classical case. The classical case gives $cost(NE)=10$ and $cost(OPT)=7.5$, while Ref.~\cite{scarpa} has managed to reduce $cost(NE)$ to $cost(OPT)$ such that $PoS=PoA=1$, we have further reduced the quantum optimal cost itself resulting in $cost(OPT)=cost(NE)=7.176$ for the case with strategy $Q$ and $cost(OPT)=cost(NE)=7.15$ for strategy $M$.
In the next section, we will conclude our paper by stating the main outcomes and giving a brief overview of our future endeavors.
\section{CONCLUSION AND PERSPECTIVE}
In this paper, we started off by giving a brief introduction to Pigou's problem and its applications. On solving the two person classical Pigou's problem, we find that the best possible outcome would be the case where traffic is divided equally between both paths which is the optimal solution. However, the Nash equilibrium is not the optimal solution and leads to increased cost. Similar to this example with vehicles or people, one can also think of Pigou's example as information network with bits moving from one node to another and therein too data experience congestion leading to increase in costs.\\
Two person Pigou's network can be generalised to a $k$ person Pigou's network. Classically, this $k$ person network gives the optimal cost, $cost(OPT)=3n/4$ where $n$ is total number of people. The benefit of analysing this situation shows itself in its quantum version which gives us better outcomes, both $cost(OPT)$ and $cost(NE)$ are much reduced as compared to the classical case.\\
In the two person/qubit quantum Pigou's network, the best possible outcome is the case with strategies $P_1$, $P_2$, $M$ with maximal entanglement giving $cost(OPT)=3/2$, $cost(NE)=7/4$ and $PoS=PoA=7/6$ where the Nash equilibrium is ($M$, $M$) and optimal solution is either ($P_1$, $P_2$) or ($P_2$, $P_1$).\\
In $k$ qubit quantum Pigou's network with strategies $P_1$, $P_2$, $M$, we obtain a pure Nash equilibrium for all $k$. The optimal solution and Nash equilibrium obtained in $k=4$ and $k=5$ cases gives minimal cost with $cost(OPT)=cost(NE)=7.15$ and $PoS=PoA=1$. Hence, the most feasible option for quantum networks is when data encoded in qubits moves between nodes using strategy $M$. We conclude that both quantum $k$ qubit Pigou's network and quantum two qubit Pigou's network employing quantum strategies and having maximal entanglement significantly reduces congestion costs.\\
In future, we will extend the quantum Pigou's network to analyse Braess's Paradox too.
\begin{widetext}
\section{Appendix}
{\scriptsize
\subsection{Quantum two person Pigou's network}
\textbf{This mathematica code is to show the calculations of cost in Table~\ref{tab:t3} for strategies ($\textit{P}_{\text{1}}$, $\textit{P}_{\text{1}}$) in quantum two qubit Pigou{'}s network, see section II.B.1.(a). One can write a similar code for other strategies.}
\begin{spacing}{1.75}
\noindent\(\\U_A=\{\{1,0\},\{0,1\}\}\rightarrow\text{Alice's strategy matrix i.e., $P_1$}\\
\textit{P2}=\{\{0,1\},\{-1,0\}\}\rightarrow\text{Strategy matrix associated with path $P_2$}\\
T' = \{\{0,0,0,i \pi/4\},\{0,0,-i \pi/4,0\},\{0,-i \pi/4,0,0\},\{i \pi/4,0,0,0\}\}\rightarrow\text{gives the value of $i\frac{\pi}{4}P_2\otimes P_2$}\\
J = \text{MatrixExp}[-T']\rightarrow\text{gives the entangling operator}\\
J=\left\{\left\{\frac{1}{\sqrt{2}},0,0,\frac{-i}{\sqrt{2}}\right\},\left\{0,\frac{1}{\sqrt{2}},\frac{i}{\sqrt{2}},0\right\},\left\{0,\frac{i}{\sqrt{2}},\frac{1}{\sqrt{2}},0\right\},\left\{\frac{-i}{\sqrt{2}},0,0,\frac{1}{\sqrt{2}}\right\}\right\}\\
J^{\dag}=\left\{\left\{\frac{1}{\sqrt{2}},0,0,\frac{i}{\sqrt{2}}\right\},\left\{0,\frac{1}{\sqrt{2}},\frac{-i}{\sqrt{2}},0\right\},\left\{0,\frac{-i}{\sqrt{2}},\frac{1}{\sqrt{2}},0\right\},\left\{\frac{i}{\sqrt{2}},0,0,\frac{1}{\sqrt{2}}\right\}\right\}\rightarrow\text{gives the disentangling operator}\\
|00\rangle = \{\{1\},\{0\},\{0\},\{0\}\}\\
\langle00|=\{\{1,0,0,0\}\}\\
|01\rangle=\{\{0\},\{1\},\{0\},\{0\}\}\\
\langle01|=\{\{0,1,0,0\}\}\\
|10\rangle=\{\{0\},\{0\},\{1\},\{0\}\}\\
\langle10|=\{\{0,0,1,0\}\}\\
|11\rangle=\{\{0\},\{0\},\{0\},\{1\}\}\\
\langle11|=\{\{0,0,0,1\}\}\\
\psi=\text{Dot}[J,|00\rangle]\rightarrow\text{Initial state of game}\\
U_B=\{\{1,0\},\{0,1\}\}\rightarrow\text{Bob's strategy matrix i.e., $P_1$}\\
M=\text{TensorProduct}[U_A,U_B]\\
M'=\{\{1,0,0,0\},\{0,1,0,0\},\{0,0,1,0\},\{0,0,0,1\}\}\rightarrow\text{gives $U_A\otimes U_B$}\\
K=\text{Dot}[M',\psi]\\
\psi_f=\text{Dot}[J^{\dag},K]\rightarrow\text{Final state of the game}\\
P_{00}=(\text{Abs}[\text{Dot}[\langle00|,\psi_f]]){}^{\wedge}2\\
P_{01}=(\text{Abs}[\text{Dot}[\langle01|,\psi_f]]){}^{\wedge}2\\
P_{10}=(\text{Abs}[\text{Dot}[\langle10|,\psi_f]]){}^{\wedge}2\\
P_{11}=(\text{Abs}[\text{Dot}[\langle11|,\psi_f]]){}^{\wedge}2\\
\$_{\text{A}}=1*P_{00}+1*P_{01}+1/2*P_{10}+1*P_{11}\\
\$_{\text{B}}=1*P_{00}+1/2*P_{01}+1*P_{10}+1*P_{11}\\\)
\noindent\makebox[\linewidth]{\rule{0.8\paperwidth}{0.4pt}}
\noindent\(P_{00}=1\\\)
\noindent\(P_{01}=0\\\)
\noindent\(P_{10}=0\\\)
\noindent\(P_{11}=0\\\)
\noindent\(\$_{\text{A}}=1\\\)
\noindent\(\$_{\text{B}}=1\)
\end{spacing}
\vspace{-0.5cm}
\subsection{Quantum \textit{k} person Pigou's network}
\textbf{This mathematica code is to show the calculations of cost in Table~\ref{tab:t6}, \textit{cost(NE)} and \textit{PoS, PoA} for strategies (\textit{M, M}) for \textit{k = 1} in \textit{k} qubit quantum Pigou{'}s network in section II.B.2. One can write a similar code for other strategies.}
\begin{spacing}{1.75}
\noindent\(\\k=1\\
n=10\\
U_A=\frac{1}{\sqrt{2}}\{\{i,1\},\{-1,-i\}\}\rightarrow\text{Alice's strategy matrix i.e., $M$}\\
\textit{P2}=\{\{0,1\},\{-1,0\}\}\rightarrow\text{Strategy matrix associated with path $P_2$}\\
T' = \{\{0,0,0,i \pi/4\},\{0,0,-i \pi/4,0\},\{0,-i \pi/4,0,0\},\{i \pi/4,0,0,0\}\}\rightarrow\text{gives the value of $i\frac{\pi}{4}P_2\otimes P_2$}\\
J = \text{MatrixExp}[-T']\rightarrow\text{gives the entangling operator}\\
J=\left\{\left\{\frac{1}{\sqrt{2}},0,0,\frac{-i}{\sqrt{2}}\right\},\left\{0,\frac{1}{\sqrt{2}},\frac{i}{\sqrt{2}},0\right\},\left\{0,\frac{i}{\sqrt{2}},\frac{1}{\sqrt{2}},0\right\},\left\{\frac{-i}{\sqrt{2}},0,0,\frac{1}{\sqrt{2}}\right\}\right\}\\
J^{\dag}=\left\{\left\{\frac{1}{\sqrt{2}},0,0,\frac{i}{\sqrt{2}}\right\},\left\{0,\frac{1}{\sqrt{2}},\frac{-i}{\sqrt{2}},0\right\},\left\{0,\frac{-i}{\sqrt{2}},\frac{1}{\sqrt{2}},0\right\},\left\{\frac{i}{\sqrt{2}},0,0,\frac{1}{\sqrt{2}}\right\}\right\}\rightarrow\text{gives the disentangling operator}\\
|00\rangle = \{\{1\},\{0\},\{0\},\{0\}\}\\
\langle00|=\{\{1,0,0,0\}\}\\
|01\rangle=\{\{0\},\{1\},\{0\},\{0\}\}\\
\langle01|=\{\{0,1,0,0\}\}\\
|10\rangle=\{\{0\},\{0\},\{1\},\{0\}\}\\
\langle10|=\{\{0,0,1,0\}\}\\
|11\rangle=\{\{0\},\{0\},\{0\},\{1\}\}\\
\langle11|=\{\{0,0,0,1\}\}\\
\psi=\text{Dot}[J,|00\rangle]\rightarrow\text{Initial state of game}\\
U_B=\frac{1}{\sqrt{2}}\{\{i,1\},\{-1,-i\}\}\rightarrow\text{Bob's strategy matrix i.e., $M$}\\
M=\text{TensorProduct}[U_A,U_B]\\
M'=\{\{-1,i,i,1\},\{-i,1,-1,-i\},\{-i,-1,1,-i\},\{1,i,i,-1\}\}\rightarrow\text{gives $U_A\otimes U_B$}\\
K=\text{Dot}[M',\psi]\\
\psi_f=\text{Dot}[J^{\dag},K]\rightarrow\text{Final state of the game}\\
P_{00}=(\text{Abs}[\text{Dot}[\langle00|,\psi_f]]){}^{\wedge}2\\
P_{01}=(\text{Abs}[\text{Dot}[\langle01|,\psi_f]]){}^{\wedge}2\\
P_{10}=(\text{Abs}[\text{Dot}[\langle10|,\psi_f]]){}^{\wedge}2\\
P_{11}=(\text{Abs}[\text{Dot}[\langle11|,\psi_f]]){}^{\wedge}2\\
\$_{\text{A}}=1*P_{00}+1*P_{01}+(k+1)/n*P_{10}+(k+2)/n*P_{11}\\
\$_{\text{B}}=1*P_{00}+(k+1)/n*P_{01}+1*P_{10}+(k+2)/n*P_{11}\\
cl=k*k/n+(n-k-2)*1\rightarrow\text{total cost for the players/qubits whose behaviour is fixed}\\
cost(NE)=\$_{\text{A}}+\$_{\text{B}}+cl\\
cost(OPT)=7.15\\
PoS=PoA=cost(NE)/cost(OPT)\\\)
\noindent\makebox[\linewidth]{\rule{0.8\paperwidth}{0.4pt}}
\noindent\(P_{00}=1/4\\\)
\noindent\(P_{01}=1/4\\\)
\noindent\(P_{10}=1/4\\\)
\noindent\(P_{11}=1/4\\\)
\noindent\(\$_{\text{A}}=5/8\\\)
\noindent\(\$_{\text{B}}=5/8\\\)
\noindent\(cost(NE)=8.35\\\)
\noindent\(PoS=PoA=1.17\)
\end{spacing}

\vspace{-0.5cm}
{\subsection{Specifics for \textit{k} person quantum Pigou's network (strategy \textit{Q})}
\textbf{This mathematica code is to show the calculations of cost(NE) and \textit{PoS}, \textit{PoA} using the data given in Table~\ref{tab:t5}, for \textit{k = 1} in \textit{k} qubit quantum Pigou{'}s network (strategy \textit{Q}) in section II.B.2. One can write a similar code for other values of \textit{k}.}
\begin{spacing}{1.75}
\noindent\(\\k=1\\
n=10\\
p1=\frac{n-k-2}{4(n-k-2)+1}\rightarrow\text{mixed strategy probability for Alice taking strategy $P_1$}\\
p2=\frac{n-k-2}{4(n-k-2)+1}\rightarrow\text{mixed strategy probability for Alice taking strategy $P_2$}\\
1-p1-p2=\frac{2(n-k-2)+1}{4(n-k-2)+1}\rightarrow\text{mixed strategy probability for Alice taking strategy $Q$}\\
q1=\frac{n-k-2}{4(n-k-2)+1}\rightarrow\text{mixed strategy probability for Bob taking strategy $P_1$}\\
q2=\frac{n-k-2}{4(n-k-2)+1}\rightarrow\text{mixed strategy probability for Bob taking strategy $P_2$}\\
1-q1-q2=\frac{2(n-k-2)+1}{4(n-k-2)+1}\rightarrow\text{mixed strategy probability for Bob taking strategy $Q$}\\
A=\{\{p1,p2,1-p1-p2\}\}\\
B=\{\{q1,q2,1-q1-q2\}\}\\
\$_{\text{A}}=\text{Dot}[A, {{1, 1, (k + 2)/10}, {(k + 1)/10, (k + 2)/10, 
   1}, {(k + 2)/10, (k + 1)/10, 1}}, \text{Transpose}[B]]\rightarrow\text{Alice's payoff calculated using the method given in Ref.~\cite{devos}.}\\
\$_{\text{B}} = \text{Dot}[A, {{1, (k + 1)/10, (k + 2)/10}, {1, (k + 2)/10, (k + 1)/
     10}, {(k + 2)/10, 1, 1}}, \text{Transpose}[B]]\rightarrow\text{Bob's payoff calculated using the method given in Ref.~\cite{devos}.}\\
cl = k*k/10 + (10 - k - 2)*1\rightarrow\text{total cost for the players/qubits whose behaviour is fixed}\\
cost(NE)=\$_{\text{A}}+\$_{\text{B}}+cl\\
cost(OPT)=7.176\rightarrow\text{$cost(NE)$ for $k=4$ case as the optimal solution changes to this case.}\\
PoS=PoA=cost(NE)/cost(OPT)\\\)
\noindent\makebox[\linewidth]{\rule{0.8\paperwidth}{0.4pt}}
\noindent\(p1=7/29\\\)
\noindent\(p2=7/29\\\)
\noindent\(1-p1-p2=15/29\\\)
\noindent\(q1=7/29\\\)
\noindent\(q2=7/29\\\)
\noindent\(1-q1-q2=15/29\\\)
\noindent\(\$_{\text{A}}=37/58\\\)
\noindent\(\$_{\text{B}}=37/58\\\)
\noindent\(cost(NE)=8.38\\\)
\noindent\(PoS=PoA=1.17\)
\end{spacing}}
}
%% AMS-LaTeX Created with the Wolfram Language : www.wolfram.com
\end{widetext}

\end{document}